\documentclass[letterpaper,twocolumn,10pt]{article}
\usepackage{usenix,epsfig,endnotes}
\usepackage[utf8]{inputenc}
\usepackage[T1]{fontenc}
\usepackage{amsmath}
\usepackage{txfonts}
\usepackage{todonotes}
\usepackage{tikz}
\usepackage{url}
\usetikzlibrary{arrows,shapes,decorations.pathreplacing}

\newcommand\code[1]{\texttt{#1}}
\newcommand\s{\textrm{s}}
\newcommand\ms{\textrm{ms}}

\begin{document}

\title{\Large \bf User-space Multipath UDP in Mosh}
\author{
{\rm Matthieu Boutier}\\
boutier@pps.univ-paris-diderot.fr\\
\and
{\rm Juliusz Chroboczek}\\
jch@pps.univ-paris-diderot.fr\\
\and
Sorbonne Paris Cit\'e, PPS, UMR 7126, CNRS, F-75205 Paris, France
}
\date{}

\maketitle

\section*{Abstract}
In many network topologies, hosts have multiple IP addresses, and may
choose among multiple network paths by selecting the source and
destination addresses of the packets that they send.  This can happen with
multihomed hosts (hosts connected to multiple networks), or in multihomed
networks using source-specific routing \cite{boutier2014ssr}.  A number of
efforts have been made to dynamically choose between multiple addresses in
order to improve the reliability or the performance of network
applications, at the network layer, as in Shim6 \cite{garcia2010shim6}, or
at the transport layer, as in MPTCP \cite{mptcp2012}.  In this paper, we
describe our experience of implementing dynamic address selection at the
application layer within the \emph{Mosh} Mobile Shell
\cite{winstein2012mosh}.  While our work is specific to Mosh, we hope that
it is generic enough to serve as a basis for designing UDP-based multipath
applications or even more general APIs.
\section{Introduction}

Standard networking APIs are mainly designed with the implicit assumption
that a client with a single address connects to a server with a single
address.  This assumption is often incorrect: many hosts have multiple
addresses, either because they are multihomed (connected to multiple
networks) or connected to a network that is itself multihomed.

\paragraph{Multihomed hosts}

\begin{figure}[hb]
\centerline{
\begin{tikzpicture}[-, auto, node distance=3cm, thick, main
  node/.style={circle,draw,minimum width=.5cm}]
  \draw (0, 0) node (0) {WiFi};
  \draw (2, 0) node[main node] (1) {};
  \draw (4, 0) node (2) {3G};
  \path
    (0) edge node {} (1)
    (1) edge node {} (2)
    ;
\end{tikzpicture}}
\caption{A multihomed host}\label{fig:mh-host}
\end{figure}
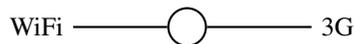

Many hosts have are connected to multiple networks (Figure~\ref{fig:mh-host}).
The most common case is that of a mobile host connected to the Internet
over multiple network technologies (e.g.\ WiFi and cellular or WiFi and
Ethernet), but this is also the case of servers with redundant
connectivity.  Since hosts do not typically participate in the routing
protocol, multihomed hosts have multiple addresses, one per network
interface.

A similar situation arises with double-stack hosts, that have both IPv4
and IPv6 addresses.  While both addresses are assigned to the same network
interface, such hosts are multihomed from the point of view of the higher
layers.

\paragraph{Source-specific routing and multihomed networks}

\begin{figure}[hb]
\centering{
\begin{tikzpicture}[-, auto, node distance=3cm, thick, main
  node/.style={circle,draw,minimum width=.5cm}, scale=.85]
  \draw (-0.4, .5) node (0) {ISP 1};
  \draw (1, .5) node[main node] (1) {};
  \draw (2.5, 1) node[main node] (2) {};
  \draw (3, 0) node[main node] (3) {};
  \draw (4, 1) node[main node] (4) {};
  \draw (4.5, 0) node[main node] (5) {};
  \draw (6, .5) node[main node] (6) {};
  \draw (7.4, .5) node (7) {ISP 2};
  \draw[dashed] (3.5, .5) ellipse(2 and 1);
  \path
    (0) edge node {} (1)
    (1) edge node {} (2)
    (2) edge node {} (3)
    (2) edge node {} (4)
    (3) edge node {} (4)
    (3) edge node {} (5)
    (4) edge node {} (5)
    (5) edge node {} (6)
    (6) edge node {} (7)
    ;
\end{tikzpicture}}
\caption{A network connected to two providers}\label{fig:mh-net}
\end{figure}
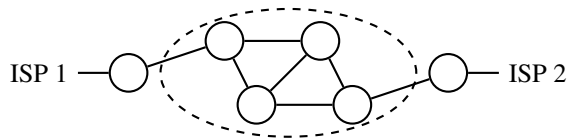

Similar to a multihomed host, a multihomed network
(Figure~\ref{fig:mh-net}) is one that is connected to the Internet over
multiple physical links, either for fault tolerance or in order to improve
throughput or reduce cost.  Classically, such networks are assigned
\emph{Provider-Independent} (PI) addresses that are announced over all
links, in which case the dynamic nature of the routing protocol
automatically provided for fault-tolerance; improvements in throughput and
reductions in cost can be achieved by careful engineering of the routing
protocol.

PI addresses need to be announced in the Internet's global ``Default-Free
Zone'' and cannot be easily aggregated; they are therefore a costly
resource.  For smaller networks, it is a natural proposition to announce
multiple \emph{Provider-Dependent} prefixes, one to each provider.  In
this case, the internal routing protocol must be able to perform
\emph{source-specific routing} (sometimes called \emph{SADR}), where
outgoing packets are routed to the right gateway depending on their
\emph{source}.  We describe our experience with source-specific routing
elsewhere \cite{boutier2014ssr}.

\subsection{Choosing addresses dynamically}

In the presence of multiple addresses, hosts must choose a pair of
a source and a destination address for every packet that they send.  The
usual approach is to make this choice at connection establishment: the set
of possible (source, destination) pairs is ordered according to a set of
static rules \cite{rfc3484}, and the first that works is used throughout
the lifetime of the connection.  This is sometimes slightly refined by
probing all of the possible pairs in parallel \cite{rfc6555}.

An obvious drawback of such a relatively static approach is that the
choice of addresses cannot be changed once a connection has been
established, with the consequence that a network hiccup will cause all
established connections to drop (or, worse, to hang).  A number of
researchers have come up with protocols that switch addresses
mid-connection; this can be done at the network layer, at the transport
layer, or, as in our work, at the application layer.

\paragraph{Network layer}
Shim6~\cite{garcia2010shim6,rfc5533} is a host-centric layer 3 shim which
provides reliability in multihomed networks.  When a connection is established
by higher layers to a remote host, Shim6 exchanges the local and remote
addresses.  If the connection is broken, Shim6 finds an alternative working path
and changes the source and destination addresses of both the outgoing and
incoming packets of the connection.  This operation is totally transparent
to the higher layers.

Shim6 detects failures based on the natural traffic of the application: if
packets are sent without response, then it is likely that a failure has
occurred.  In that case, Shim6 finds another responsive path using the
reachability protocol (REAP)~\cite{rfc5534}.

REAP builds a list of all possible paths, and probes them to find one that
works.  REAP probes can be fairly heavyweight, since they contain
information about the other probes sent and received, allowing REAP to detect
and benefit from unidirectional paths.  To limit the overhead of probes,
REAP sorts the flows with a number of heuristics before probing them with
an exponential back-off interval between two probes.  Naderi et
al.~\cite{naderi2014probing} remark that having more aggressive probing is
desirable in practice, since it achieves fastest convergence while
the overhead will likely be less than the broken connection's traffic.

\paragraph{Transport layer}

Multiple source and destination addresses can be chosen by the transport
protocol, as is the case in Multipath TCP (MPTCP)~\cite{mptcp2012} and
SCTP~\cite{rfc4960}.

MPTCP is a reliable flow-oriented TCP-compatible protocol which establishes as
many alternative paths as possible.  These paths are called \emph{subflows}, and
provide both reliability and better performance: MPTCP is able to balance
traffic between its subflows, without performance losses due to slow flows.
Subflows are continuously estimated by regular TCP messages, which act as
probes.

SCTP is a reliable, message-oriented, connection-oriented protocol,
designed to achieve reliability in multihomed networks.  When a SCTP
connection \emph{(association)} is established, a \emph{primary path} is
selected by each end-point and used to send packets to the other one.
A number of mechanisms allow the end-points to exchange their addresses
and build alternate paths.  These paths are regularly probed at at fixed
intervals \emph{(heartbeats)} to check their reliability.  When a failure
is detected on the primary path, one of the working alternate paths is
selected to be the new primary path.  The protocol definition gives no
guidelines about the strategy to be used to switch to a better path when
the primary path is active but less efficient than other paths.

\paragraph{Application layer}
A server's addresses are usually stored within the DNS, and the client will try
them all, either in turn \cite{rfc3484} or in parallel \cite{rfc6555}.  Once a
flow is established, it is no longer possible to change the source and
destination addresses --- from the user's point of view, all connections are
broken whenever a link outage forces a change of address.

A TCP connection is bound to a pair of addresses at connection
establishment; however, a number of modern applications are based upon UDP
and implement their own scheduling and retransmission strategies.  This is
notably the case of VoIP applications (SIP, Skype, etc.), of some recent
peer-to-peer file transfer applications \cite{norbert2012utp}, and of
\emph{Mosh}, the \emph{Mobile Shell}, an SSH replacement designed for
lossy and high-latency links \cite{winstein2012mosh}.

Since standard Mosh doesn't implement full multipath, it is unable to
survive certain classes of topology changes.  In the rest of this paper,
we describe our implementation of multipath for Mosh, and our experiences
with our implementation.

\section{The original Mobile Shell}\label{sec:stdmosh}

The Mobile Shell is a lightweight and responsive remote shell.  It differs from
SSH mainly by two aspects.  On the one hand, it predicts what should display the
terminal while waiting for a server confirmation, which increase the user
experience on slow links, and on the other hand, it uses the UDP transport
protocol with an algorithm limiting the impact of packet loss and reordering.

Mosh is composed of multiple layers: the front-end layer, the transport layer
and the network layer.  The front-end layer is in charge of interaction with the
user, client-side, and with the remote host, server-side.  It includes the
prediction algorithm, and communicates with the server through the Mosh
transport layer.

The transport layer is in charge of keeping the states of the two peers in
sync and of data fragmentation.  It is also its responsibility to deal
with packet loss and reordering: essentially, it maintains the knowledge of the
last acknowledged state, and tries to send deltas from that state to the
current local state.  If a packet is lost, its content will (probably) be
included in the next packet, avoiding the need for a retransmission.
Reordering is also not a problem: as newer packets have also the
information bring by the older, the older has just to be ignored.

The network layer is in charge to receive and send packets, estimate the Maximum
Transmission Unit (MTU), Round Trip Time (RTT) and Retransmission TimeOut (RTO),
encrypt and decrypt packets, and keep connection active, even in the presence of
NAT.  Our modifications to enable multipath are confined to the network layer.

\paragraph{Mosh network protocol}

The Mosh network protocol header is described in Figure \ref{fig:format},
and consists of 12 octets (96 bits) The first 64 bits contain
a cryptographic nonce, a flag indicating the direction of the flow (client
to server, or server to client), and the message sequence number (seqno).
The rest of the header consists of two timestamp fields used to measure the RTT.

\begin{figure}
\centerline{
\begin{tabular}{|c|c|@{}l}
  \cline{1-2}
  \begin{tabular}{@{}c@{}}
    1  bit \\
    63 bits \\
  \end{tabular} &
  \begin{tabular}{@{}c@{}}
    direction \\
    sequence number \\
  \end{tabular} &
  $\left.\vphantom{\begin{array}{c}.\\.\end{array}}\right\}
  \mathrm{64\ bits\ nonce}$ \\
  \cline{1-2}
  16 bits & timestamp \\
  \cline{1-2}
  16 bits & timestamp reply \\
  \cline{1-2}
\end{tabular}}
\caption{Original Mosh packet format}
\label{fig:format}
\end{figure}

\paragraph{RTT measurement}

When a packet is about to be sent, the timestamp field is filled with the host's
current timestamp, and the timestamp reply field is set to the last timestamp
received from the remote peer plus the time elapsed since the packet bringing
that timestamp was received.  The RTT is computed as the difference of the
current timestamp and the timestamp reply field of each packet; and is averaged
by an exponential mean to compute the smoothed RTT (SRTT).

\begin{figure}
\centerline{
\begin{tikzpicture}[-, auto, thick, main node/.style={draw}]
  \draw (0, 6.5) node[above] {client} -- (0, 2.5);
  \draw (3, 6.5) node[above] {server} -- (3, 2.5);
  \draw[->] (0, 6) node[left] {$t_0$} -- (3, 5.5) node[above right] {$u_0$}
  node[sloped,midway,above] {$(0, s, t_0, -1)$};
  \draw (3, 5) node[right] {$\Delta$};
  \draw[->] (3, 4.5) node[below right] {$u_1$} -- (0, 4) node[above left] {$t_1$}
  node[sloped,midway,above] {$(1, s', u_1, t_0 + \Delta)$};
  \draw (0, 3.75) node[left] {$\Delta'$};
  \draw[->] (0, 3.5) node[below left] {$t_2$} -- (3, 3) node[above right] {$u_0$}
  node[sloped,midway,above] {$(0, s+1, t_2, u_1 + \Delta')$};
\end{tikzpicture}
}
\caption{A Mosh network layer exchange}
\label{fig:timestamp}
\end{figure}
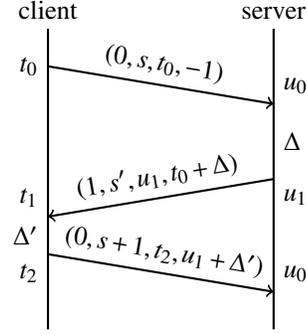

Figure~\ref{fig:timestamp} illustrates this process: the client sends
a packet destined to the server ($\mathrm{destination} = 0$), with seqno
$s$, timestamp $t_0$, and timestamp reply $-1$, because it has not yet
received a timestamp from the server.  The server receives the packet at
time $u_0$, and stores both $t_0$ and $u_0$ as the last timestamps
received from the client.  When it next sends data to the client, at time
$u_1$, it fills the packet with the right destination field
($\mathrm{destination} = 1$), its seqno $s'$, the timestamp $u_1$, and the
timestamp reply $t_0' = t_0 + \Delta = t_0 + (u_1 - u_0)$.  On reception,
the client computes the value of the RTT as $\mathit{RTT} = t_1 - t_0'$,
and stores the two timestamps for future messages.

In order to keep accurate measurements, only in-order messages are considered,
i.e.\ messages with a strictly increasing sequence number.  The payload of
out-of-order packets is still transmitted to the transport layer, but the
control information is ignored.

\paragraph{Connection timeout}
As Mosh uses UDP, an unreliable transport protocol, it must determine
when the connection has been lost.  Mosh uses the standard RTO computation, as
defined in TCP: $\mathit{RTO} = \mathit{SRTT} + 4 \times \mathit{SRTTVAR}$.  The
timeout value is not used by the network layer, but by the higher layers,
so that data are sent at regular intervals to continuously monitor the
connection, and to inform the user that the connection has been lost.

\paragraph{Connection loss}
A connection can be lost due to many causes: network failure, link
removal, changing IP addresses, or NAT timeouts.  Mosh makes a best
effort to keep the connection active, and will never give up without an
explicit request from the user.

If the connection times out, data packets are still sent to the server, but at
larger intervals.  When the network offers connectivity again, the connection
recovers almost immediately.

Mosh resists to IP address changes because by using the \code{sendto}
system call with no explicit source address, leaving the responsibility of
choosing the source address to the operating system.  It is therefore
possible to use the same Mosh session from multiple addresses (e.g. after
moving a laptop).

NATs sometimes time out active connections.  As NATs associate
a connection to the 5-tuple (transport protocol, IP src, IP dst, port src,
port dst), a Mosh client switches ports regularly when it has no sign of
the server.  This is done by creating and using a new socket to send data.
Mosh retains a small number of older sockets, in case a packet from the
server arrives for a previously used port.

\paragraph{Limitations of Mosh}
In some cases, the mechanisms implemented in Mosh are not sufficient to
keep a connection, or simply to choose the best connection among multiple
possible ones.  In a multi-address environment, a Mosh client may have
multiple paths to reach the destination; since Mosh lets the system choose
the source address, it will not switch to a better path or change protocol
familites.

\section{Multipath Mosh}\label{sec:mpmosh}
There are potentially as many different paths as pairs $(s,d)$ of
addresses, where $s$ is a client address and $d$ a server address.  Our
implementation of a multipath Mosh continually estimates the latency of
each of these paths and selects the best one.  We call \emph{flow} the
structure representing a path and its characteristics; this corresponds to
MPTCP's subflows or SCTP's paths.

We have added a \emph{flag} field to the network layer protocol in order
to have more types of messages.  We use it to differentiate data, probes,
and address gathering.  Protocol details are described in
Section~\ref{sec:details}.

\subsection{Address gathering}
In order to build all possible flows, the client must gather both its
local addresses and the server's addresses.  Both the client and the
server have access to their local addresses through standard APIs (such as
\code{getifaddrs}) provided by the system.  At bootstrap, the client has
a reachable address of the server: in combination with its local
addresses, it builds an initial set of flows.

\paragraph{Gathering server's addresses}
The client asks the server for its addresses at regular intervals.  In
particular, it sends an address request at bootstrap to quickly have a complete
set of flows.  Both the request and the reply messages are tagged with the
\code{ADDR\_FLAG} flag bit.

On reception of a message with the \code{ADDR\_FLAG} set, the server answers
with the set of its local addresses.  The message bringing the answer is also
tagged with the \code{ADDR\_FLAG}, and may not contain other data than the
server's addresses.

On reception of a message with the \code{ADDR\_FLAG} set, the client parses the
message and extracts the server's local addresses.  It then completes the set of
flows it has to the server.  Existing flows are not modified.

\paragraph{Address pair filtering}
Some pairs of addresses will not lead to a valid flow: for example, an
IPv4 source address cannot be used with an IPv6 destination address.
Filtering such incompatible combinations can reduce the number of flows
being probed.  Of course, determining if combination is valid or not is
not always obvious, and it is more desirable to consider too many flows
rather than filter out good ones.

We only do very limited filtering in our implementation, such as verifying
that the two addresses of the same pair have the same address family, are
both loopback addresses or not, and are both link-local or not.  While
more discrimination would be possible, we prefer to keep a more general
implementation, the overhead being tolerable (see Section~\ref{sec:overhead}).

\subsection{Sending packets on a flow}
To send a datagram on a given flow, we must specify both its source and
destination IP addresses.  There are three possible implementations:

\begin{itemize}
\item using one socket per flow, each bound both to a local address (with
  \code{bind}) and to a remote address (with \code{connect}).  Sending and
  receiving messages on a particular flow is then done with the \code{send} and
  \code{recv} system calls: the socket parameters indicate without ambiguity
  which are the local and remote addresses of both outgoing and incoming
  packets.
\item using one socket per local address, in combination with the \code{sendto}
  and \code{recvfrom} system calls, which respectively set the destination of
  outgoing packets, and retrieve the remote peer address of incoming packets.
  The local address is specified and retrieved as previously, by the socket.
\item using one socket per network-layer protocol (IPv4, IPv6), with the
  \code{sendmsg} and \code{recvmsg} system calls which allow to specify and
  retrieve both the source and the destination address of each packet.
\end{itemize}

After a number of experiments, we have decided to use the third one: it
results in a much simpler implementation, especially in the presence of
port hopping (see above).

More precisely, messages are sent with the \code{sendmsg} system call,
with the \code{msg\_name} and \code{msg\_namelen} fields providing the
destination address.  The source address is provided by a control message
(\code{msg\_control} and \code{msg\_controllen}).  In IPv4, we use
\code{IP\_PKTINFO} and \code{IP\_SENDSRCADDR}, depending on the operating
system; IPv6 provides the portable API \code{IPV6\_PKTINFO}.  (We have
found out that this only works if the socket is explictly bound to the
wildcard address --- on some systems, using an unbound socket leads to
a kernel panic.)

We did consider using a single hybrid (IPv6 and IPv4) socket.
Unfortunately, hybrid sockets' options are a small subset of plain IPv6
options, not sufficient for our needs.

\subsection{Flow selection}
Mosh is an interactive lightweight protocol: the metric we want to optimize is
clearly the RTT, and load balancing is not in our scope.  To measure the RTT,
very small messages, called probes, are regularly sent on flows to estimate
their latency.  Probe exchanges are only initiated by the client: the server
only response to probes.

We consider that paths are bidirectional and symmetric, so the client and
the server use the same flow at any one given time.  This flow is selected
by the client, which chooses the flow with lowest RTT, while the server
merely selects the flow of the latest in-order data packet received to
send data back.

We didn't implement any mechanism to limit instability, which is not
a problem in the case of Mosh.  Indeed, having instability with two paths
of roughly same RTT will at worst cause packet reordering, which is not
a problem for the higher layers of Mosh.  Our tests with flows of similar
RTT exhibited good behaviour, but more investigation in this area may be
needed.

\subsection{Probing flows}

Probing flows consists in periodically sending small messages, known as
\emph{probes}, that carry enough information to estimate the flow's
RTT. The interval between two probes is determined dynamically: waiting
for a flow RTT update should be of the same order of magnitude as the RTT
itself; in effect, faster paths will be probed more often than slower
ones.  Beyond a certain frequency, however, sending more probes no longer
improves the user experience.  For that reason, our implementation never
sends more than one probe every $500\,{ms}$.

An additional issue is that of non-functional flows: since no probes will be
received, the RTT will not be naturally increased, while a single packet
loss should not be considered as a broken path.  We maintain an
$\mathit{idle\_time}$ field in the flow structure which represents the
time during which a flow has not been responsive.  The SRTT value of an
idle flow is unchanged, but the value used for choosing flows is the sum
of the SRTT and the idle time.

A packet loss, or lack of response, is detected using RTO.  We use the
standard TCP's RTO computation, also taking the idle time into account:

\[ \mathit{RTO} := (\mathit{SRTT} + \mathit{idle\_time}) + 4 \times \mathit{SRTTVAR} \]

When a RTO has elapsed with no response from the server, the client immediately
sends another probe, even if the next probe for this flow was delayed, and the
idle time is increased by one RTO.  Increasing by no more than one RTO is
important, even if more time elapsed since the last probe sending: the event
loop may give us back control after a lot of time, leading to a significant
over-estimation of idle time on a simple packet loss.  We observed delays of
several seconds, leading the mosh client to choose a worst flow.  On the other
hand, increasing by one RTO is enough: since the computation of the RTO takes
the idle time into account, it is the same behaviour on idle flows as doubling
the RTT each time:

\[ \mathit{idle\_time} := \mathit{idle\_time} + \mathit{RTO} \]

Mosh uses delayed acknowledgements.  To compensate, the Mosh network layer
receives the maximum delay interval from the higher layers, and computes
a delayed RTO (dRTO).  The dRTO is used to compute the probing interval
and the actual timeout of packet acknowledgements, but we keep the RTO
value to increase the idle time: the delays induced by the delayed
acknowledgements are not representative of the real idle time.  This leads
to the following equations:

\begin{align*}
  \mathit{dRTO} & := \mathit{RTO} + \mathit{max\_delack} \\
  \mathit{probe\_interval} & := \max(\mathit{dRTO}, 500\,{ms})
\end{align*}

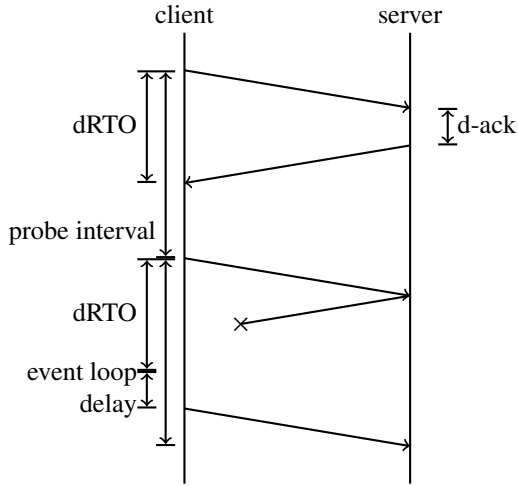
\begin{figure}
\begin{tikzpicture}[-, auto, thick, main node/.style={draw}]
  \draw (0, 8) node[above] {client} -- (0, 2);
  \draw (3, 8) node[above] {server} -- (3, 2);

  \draw[->] (0, 7.5) -- (3, 7);
  \draw[->] (3, 6.5) -- (0, 6);
  \draw[|<->|] (-0.5, 7.5)
  -- (-0.5, 6) node[midway, left] {
    \begin{tabular}{@{}r@{}}dRTO\end{tabular}};

  \draw[|<->|] (3.5, 7)
  -- (3.5, 6.5) node[midway, right] {
    \begin{tabular}{@{}l@{}}d-ack\end{tabular}};

  \draw[|<->|] (-0.25, 7.5)
  -- (-0.25, 5) node[above left] {
    \begin{tabular}{@{}r@{}}probe interval\end{tabular}};

  \draw[->] (0, 5) -- (3, 4.5);
  \draw[] (3, 4.5) -- (3/4, 4 + .5/4) node[] {$\times$};
  \draw[|<->|] (-0.5, 5)
  -- (-0.5, 3.5) node[midway, left] {
    \begin{tabular}{@{}r@{}}dRTO\end{tabular}};
  \draw[|<->|] (-0.5, 3.5)
  -- (-0.5, 3) node[midway, left] {
    \begin{tabular}{@{}r@{}}event loop\\ delay\end{tabular}};
  \draw[|<->|] (-0.25, 5)
  -- (-0.25, 2.5) node[left] {
    \begin{tabular}{@{}r@{}}\end{tabular}};
  \draw[->] (0, 3) -- (3, 2.5);
\end{tikzpicture}
\caption{Probe scheduling}
\label{fig:probing}
\end{figure}

An example of the whole process is described in figure~\ref{fig:probing}:
the client sends a probe to the server, and schedules the next probe.  The
server receives the probe, and delays the acknowledgement before sending
back a probe.  The client receives the acknowledgement in dRTO, so it
waits for the end of the probe interval to send again a probe to the
server.  This time, the server decides to send back the probe immediately,
but the packet is lost.  The client notices the loss when the event loop
next yields control: it adds RTO to the flow's idle time, and immediately
sends back a probe, without waiting for a whole probe interval.

\subsection{Probe overhead} \label{sec:overhead}
Our approach will lead to permanent overhead in the network, with regular 
probes on each flow.  As the number of flows is the product of the
number of local addresses and the number of remote addresses, it might in
principle grow to fairly large values.

Figure~\ref{fig:overhead} shows the overhead induced by one flow,
depending on the probe interval.  These values are real-world measurements
(obtained using \emph{wireshark}).  One probe is contained in an Ethernet
frame of less than 90 octets.  An idle flow, having a probe interval
higher than 10 seconds, will lead to an overhead of less than
$10\,\mathrm{B}/\s$, while an active and efficient flow will lead to an
overhead of less than $200\,\mathrm{B}/\s$.

Considering a client and a server having 4 (compatible) addresses each, 16
flows will be built.  If each of these flows has latency better than
$500\,\ms$, we will have an overhead of less that $3\,\mathrm{kB}/\s$ for
the whole network.

\begin{figure}
\begin{tabular}{|c||c|c|c|c|c|c|c|c|}
  \hline
  Probe interval ($\ms$)    & 100 & 500 & 1000 & 10000 \\ \hline
  IPv6 overhead ($\mathrm{B}/\s$)      & 900 & 180 & 90   & 9     \\ \hline
  IPv4 overhead ($\mathrm{B}/\s$)      & 720 & 144 & 72   & 7.2   \\ \hline
\end{tabular}
\caption{Ethernet probing frame overhead of one flow.}
\label{fig:overhead}
\end{figure}

\section{Protocol details}\label{sec:details}
In this section, we describe the modifications made to the Mosh network
protocol.  The new packet header is represented in
Figure~\ref{fig:newformat}.  There are two main modifications: the
identification of flows by a flow ID, and the presence of a \code{flags}
field.

\paragraph{Flow ID}
In standard mosh, the cryptographic nonce is constituted of the direction
and the sequence number.  The direction indicates if the message was sent
by the client (with a value of 0), or by the server (value of 1), while
the sequence number is used by the received to determine whether the
packet is the most recent one.  As the nonce has double usage, and needs
to be unique, mosh gives up when the sequence number overflows (wraps
around to 0).  Since it is a 63-bit value, this is unlikely to be
a serious problem.

We use a separate flow ID, allocated by the client, to identify flows, and
sequence numbers increase independently for each flow.  Both the flow ID
and sequence number are encoded in the nonce, and the server and the
client must retain flow IDs.  The client reuses existing flow IDs when
new pairs of addresses become possible: both the client and the server will
have a limited amount of flow's structure in memory, equal to the maximum
number of simultaneous flows encountered during the session.

\begin{figure}
\centerline{
\begin{tabular}{|c|c|l}
  \cline{1-2}
  \begin{tabular}{@{}c@{}}
    1  bit \\
    15 bits \\
    48 bits \\
  \end{tabular} &
  \begin{tabular}{@{}c@{}}
    direction \\
    flow ID \\
    sequence number \\
  \end{tabular} &
  $\left.\vphantom{\begin{array}{c}.\\.\\.\end{array}}\right\}
  \mathrm{64\ bits\ nonce}$ \\
  \cline{1-2}
  16 bits & flags \\
  \cline{1-2}
  16 bits & timestamp \\
  \cline{1-2}
  16 bits & timestamp reply \\
  \cline{1-2}
\end{tabular}}
\caption{Multipath Mosh packet format}
\label{fig:newformat}
\end{figure}

\paragraph{Flags}
The Flags field is used to extend the network layer protocol.  Two flags
are defined: the \code{PROBE\_FLAG} and the \code{ADDR\_FLAG}.  Currently,
the flags serve as a message type: either none or exactly one of the two
flags will be set, identifying the type of message.

Because the main objective of a probe is to be lightweight, a probe does not
carry any extra data from higher layers, and therefore is exactly 14 bytes long.

An address request message, i.e.\ a message with \code{ADDR\_FLAG} set and
originated by the client, doesn't carry any extra-information, like a probe.  An
address response message contains the list of the server's addresses, in the
format described in figure~\ref{fig:addr-format}: 1 octet for the sub-message
length (in octets), 1 octet for the address family, 2 octets for the port
number, and the rest for the address (4 octets for IPv4 addresses, and 16 octets
for IPv6 addresses).

\begin{figure}
\centerline{\begin{tabular}{|p{1cm}|p{1cm}|p{1cm}|p{1.5cm}|}
    \hline
    \centering length & \centering family &
    \centering port & \centering address \tabularnewline \hline
\end{tabular}}
\centerline{\begin{tabular}{p{1cm} p{1cm} p{1cm} p{1.5cm}}
    \centering 1      & \centering 1      &
    \centering 2    & \centering length - 3
\end{tabular}}
\caption{Address sub-messages format}
\label{fig:addr-format}
\end{figure}

Since an address request is always sent at the beginning of the
connection, it may in principle be possible for a malicious program to
send duplicates of a message containing an address request; since the
server's reply is much larger than the request, this could be used for an
amplification attack.  For that reason, we ignore address request packets
unless they are in-order.

\section{Results}\label{sec:results}

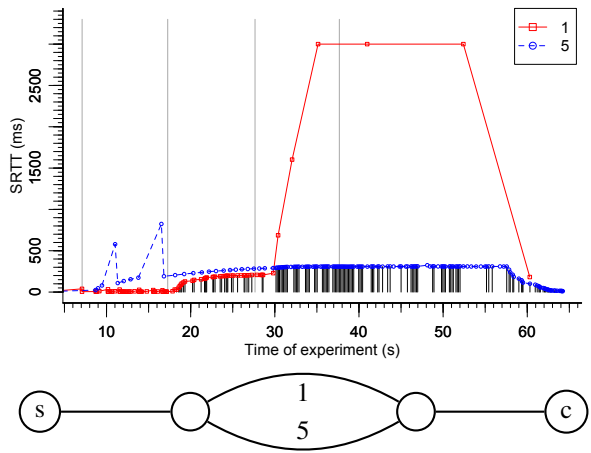
\begin{figure}
  \includegraphics[width=\linewidth]{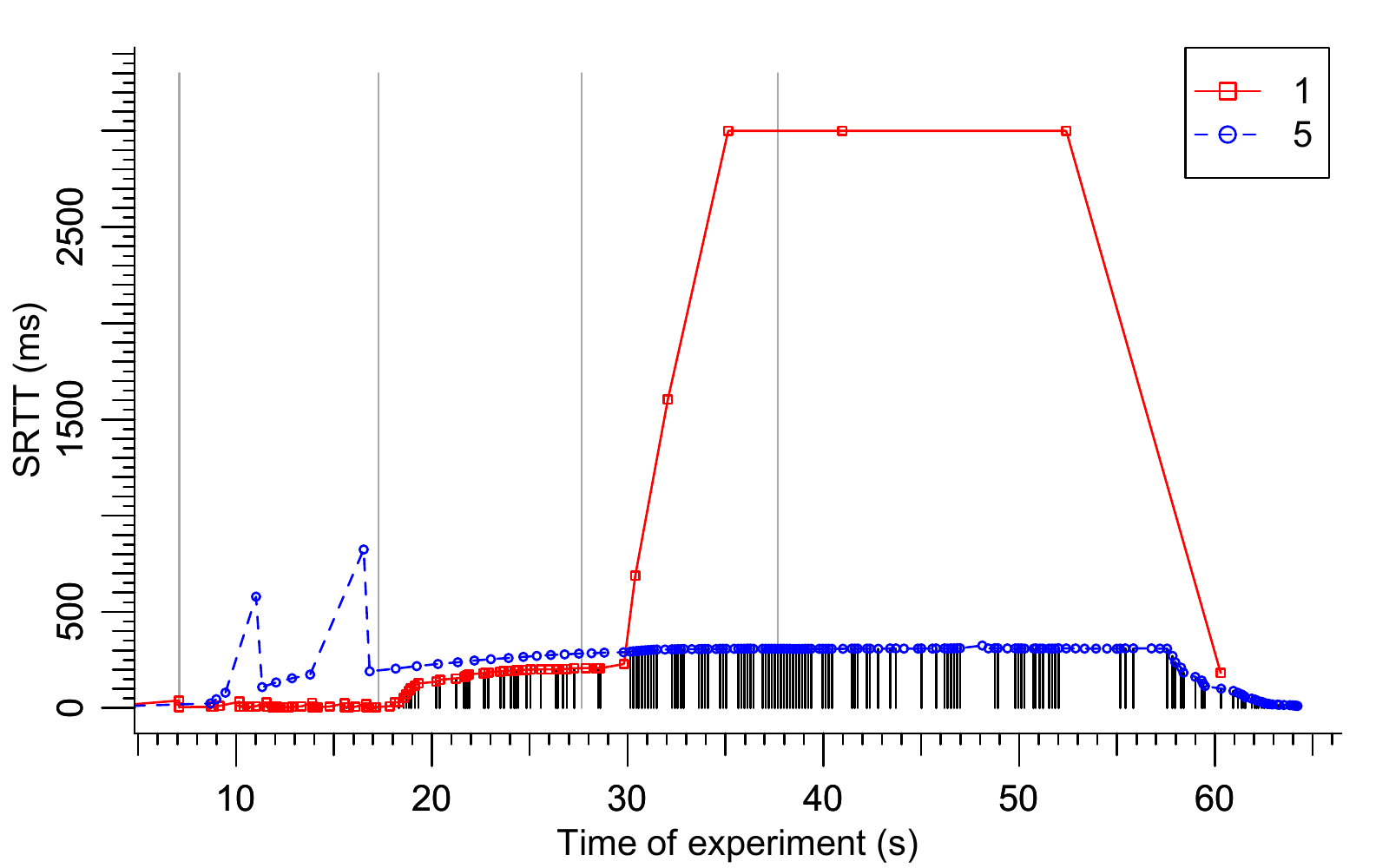}

\centerline{\begin{tikzpicture}[-, auto, node distance=3cm, thick, main
  node/.style={circle,draw,minimum width=.5cm}]
  \draw (0, 1) node[main node] (0) {s};
  \draw (2, 1) node[main node] (1) {};
  \draw (5, 1) node[main node] (2) {};
  \draw (7, 1) node[main node] (3) {c};
  \draw (1) to[bend left] node[below]{1} (2);
  \draw (1) to[bend right] node[above]{5} (2);
  \path
    (0) edge (1)
    (2) edge (3)
    ;
\end{tikzpicture}}
\caption{Testing mp-mosh on delayed or broken paths}
\label{fig:results}
\end{figure}

We tested our implementation in a simple testbed illustrated in
figure~\ref{fig:results}: the client and the server are connected to two
routers connected by two distinct paths.  We use the Linux \emph{netem}
module to simulate variable delay and loss ratios on our links, and unplug
an Ethernet cable to simulate a connection loss with no address changes.
The routing is assured by the Babel routing protocol~\cite{rfc6126}.

In this graph, we only use two flows, and we bound the higher plots to
3 seconds, both for readability purposes.  Each of the black segments at
the bottom of the graph represents a sent packet, and indicates which
flow has been chosen to send the data.  The two curves indicate the RTT
values computed by Mosh.

Initially, both paths have negligible delay.  At time $7\,\s$, we add
$300\,\ms$ delay on (the path taken by) flow 5: Mosh's estimate converges
slowly to that value while data are sent on flow 1.  This paths happened
to experiment two isolated packet losses, which increased the idle time,
but didn't disturb the SRTT value computation.

At time $17\,\s$, we add $200\,\ms$ on flow 1: its SRTT computation grows
quickly to the expected value, and data continue taking this flow, which
still has the lower latency.  The convergence of the RTT is faster, since
flow 1 received a larger number of packets.  Only a small number of probes
were sent on flow 5.

At time $28\,\s$, we unplug the Ethernet cable used by flow 1.  The first
data packets are sent on flow 5 roughly one second later, and the RTT of
flow 1 grows without bound (bounded at $3\,\s$ on the figure).

Finally, at time $38\,\s$, we plug back the Ethernet cable.  This time,
Mosh takes around 21 seconds to recover, which is the time needed for the
lower layers (notably routing) to recover, and to the probing interval
which has already reached its maximum value ($10\,\s$ in our implementation).

\section{Limitations and future work}\label{sec:limitations}
Mosh doesn't consume much bandwidth.  Contrary to MPTCP, we didn't need to
balance the load on multiple flows, but only choose the most responsive.
As the Mosh-transport layer is resilient to packet's reordering, it would
be possible to split traffic among multiple flows; we have not investigated
this possibility.

Our implementation makes no efforts to avoid instability when multiple
flows have similar latency; this did not appear to be a problem in our
experiments.  In high-throughput scenarios, Mosh may fragment messages
into multiple UDP datagrams; in that case, Mosh only buffers a single
message; if two different fragmented messages overlap, both might be
discarded.  Further experimentation is necessary in order to determine if
this is a problem in practice, and whether it is more desirable to
increase the amount of buffering done by Mosh, or to limit the amount of
instability, which might in turn decrease our algorithm's responsivity to
link outages (and perhaps the amount of natural load balancing due to
instability.)

Mosh is an application designed to have a mobile client and a fixed
server.  Client and server thus have strongly asymmetric roles.  We have
used this property, putting all the intelligence on the client side with
no choice of the path on the server side.  In other applications, both the
client and the server might be mobile, which would require smarter
server-side algorithms.

Our implementation assumes that the links are symmetric, both in
reachability and in delay.  Asymmetric protocols such as REAP require
larger probes, since each probe needs to carry information about the other
flows.  It is not clear to us whether this is worthwile in real-world
topologies, and whether the overhead can be somehow reduced.

\section{Conclusion}\label{sec:conclusion}
We have designed and implemented a multipath version of mosh, using the
assumption that using different source and destination addresses leads
to different paths.  We use active and continuous probing to estimate the
RTT of the flows induced by these paths, while having techniques to (i)
limit and adapt the number of probes depending on the performances of the
flows, (ii) quickly discriminate idle flows while not being affected by
occasional packet loss, and (iii) deal with external time delays to allow
delayed acknowledgements and event loop integration.

Our implementation achieves fast re-convergence, small overhead, and does
not interrupt the event-loop to generate extra traffic.  Our modifications
are contained in the Mosh network layer, built as a separate C++ library:
it could in principle be used by any UDP-based application that would
benefit from multiple paths.

\section{Available software}
Our multipath mosh implementation is available at:
\begin{center}
{\tt http://github.com/boutier/mosh}\\
\end{center}

\footnotesize{}


\end{document}